# Element abundances in solar energetic particles: two physical processes, two abundance patterns

**Donald V. Reames**

Institute of Physical Science and Applications, University of Maryland, College Park, MD 20742-2431 USA

E-mail: dvreames@umd.edu

**Abstract**. Abundances of elements comprising solar energetic particles (SEPs) come with two very different patterns. Historically called "impulsive" and "gradual" events, they have been studied for 40 years, 20 years by the *Wind* spacecraft. *Gradual* SEP events measure coronal abundances. They are produced when shock waves, driven by coronal mass ejections (CMEs), accelerate the ambient coronal plasma; we discuss the average abundances of 21 elements that differ from corresponding solar photospheric abundances by a well-known dependence on the first ionization potential (FIP) of the element. However, ratios such as Fe/C are fractionated by differential scattering, which varies with the mass-to-charge ratio $A/Q$ of the ion, during transport out from the shock. This makes Fe/C vary in space and time, but averaging over the spatial distribution using many SEP events can recover the FIP dependence of the source coronal abundances. The smaller *impulsive* ("$^3$He-rich") SEP events are associated with magnetic reconnection involving open field lines from solar flares or jets that also eject plasma to produce accompanying CMEs. These events produce striking heavy-element abundance enhancements, relative to coronal abundances, by an average factor of 3 at Ne, 9 at Fe, and 900 for elements with $76 \leq Z \leq 82$. This is a strong, power-law dependence on $A/Q$ with a ~3.6 power when $Q$ values are determined at coronal temperatures near 3 MK. Small individual SEP events with the steepest enhancements (~6$^{th}$ power of $A/Q$), from ~2.5 MK plasma, are associated with B- and C-class X-ray flares, and with narrow (<100°) CMEs. Enhancements in $^3$He/$^4$He can be as large as those in heavy elements but are uncorrelated with them. However, events with $^3$He/$^4$He $\geq 0.1$ are even more strongly associated with narrow, slow CMEs, cooler coronal plasma, and smaller X-ray flares. The impulsive SEP events *do not* come from hot flare plasma; they are accelerated early and/or on adjacent open field lines.

## 1. Introduction

Abundances of the chemical elements have shown a remarkable variation among the populations of energetic particles seen throughout the heliosphere [1]. This fractionation has been useful in identifying the sources of the particles and the physical mechanisms that have acted upon them. Abundances of the elements Li, Be, and B produced by nuclear fragmentation of C, N, and O during their traversal of interstellar H provided an early measure of the age of the galactic cosmic rays. Nearly pure energetic H, magnetically trapped in the Earth's inner radiation belt, was produced by the decay of neutrons generated in nuclear reactions of cosmic rays with the Earth's atmosphere. The dominant abundances of energetic S and O in the Jovian magnetosphere suggested their origin as $SO_2$ emitted from the volcanoes of the Jovian moon Io. The "anomalous cosmic rays" are dominated by elements with a first ionization potential (FIP) *above* 10 eV; these ions were neutral in interstellar space so that they easily traversed the heliospheric magnetic fields to be photoionized near the Sun, picked up by the solar wind, and carried outward to be accelerated in the outer heliosphere.

Solar energetic particles (SEPs) provide us with two particle populations (see review [2]) as distinct as those above. In the "gradual" SEP events, shock waves, driven out from the Sun by coronal mass ejections (CMEs), sample the coronal abundances of elements beginning at 2–3 solar radii. Ever since the review of Meyer in 1985 [3] it has been known that these SEP-derived coronal abundances, relative to solar photospheric abundances, show a pattern in which high-FIP (>10 eV) elements are suppressed by a factor of about 4 relative to low-FIP elements. This occurs because high-FIP elements are neutral in the photosphere while low-FIP elements are ionized. Ions are more easily transported up into the corona than neutral atoms by the action of magnetic fields and Alfvén waves, as described, for example, by Laming [4].

Impulsive SEP events involve magnetic reconnection in solar flares and jets [5] which has been found to produce strong, 1000-fold enhancements of elements, relative to coronal abundances, with high values of mass-to-charge ratio $A/Q$, both observationally [6] and theoretically [7]. Historically, these impulsive SEP events have also shown 1000-fold enhancements of $^3$He/$^4$He, relative to the that in the corona or solar wind, that are associated with streaming electrons and type III radio bursts [8]. Enhancements of heavy elements and of $^3$He are uncorrelated, suggesting a different mechanism, involving resonant wave-particle interactions, for the $^3$He enhancements [9].

In this paper we review recent studies of the element abundances in the two classes of SEP events.

## 2. Element abundance measurements

The measurement of abundances of SEPs amounts to counting of individual ions of various identified species in a particular velocity interval. Typical particle observations during SEP events at ~2–20 MeV amu$^{-1}$ are shown in Figure 1. Each ion is a point in an energy-loss *vs.* residual-energy plane. Instruments are calibrated before flight with beams of accelerated ions at known energy, including beams of C, O, Fe, Ag, and Au ions.

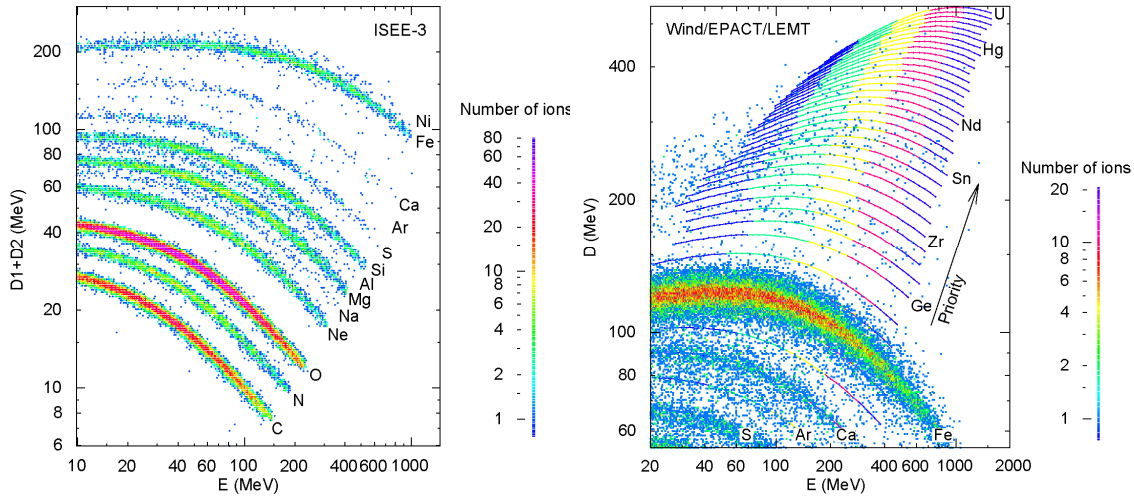

**Figure 1**. The left panel shows typical resolution of elements in large gradual SEP events [10]. The right panel shows the region above Fe where only bands of elements, comparable in width to the Fe band, are resolved [11].

## 3. Distinguishing impulsive and gradual SEP events

Ever since the discovery of $^3$He-rich events, 44 years ago (see [2]), it has been clear that more than one physical mechanism is responsible for molding SEP abundances. While the average abundances of each event class are quite clear, event-to-event variations within a class can blur some of the abundances. In addition, shock waves can reaccelerate suprathermal ions left over from previous impulsive SEP events [12]. Nevertheless, the abundances of selected elements can be used to classify sufficiently intense SEP events. The left panel in Figure 2 shows statistical separation of gradual and impulsive abundance periods in a distribution of Ne/O *vs.* Fe/O for all 8-hr averages observed in ~19 years [11]. The right panel of Figure 2 shows the time evolution of two SEP events, impulsive and gradual, showing intensities (lower right) and normalized abundances (upper right).

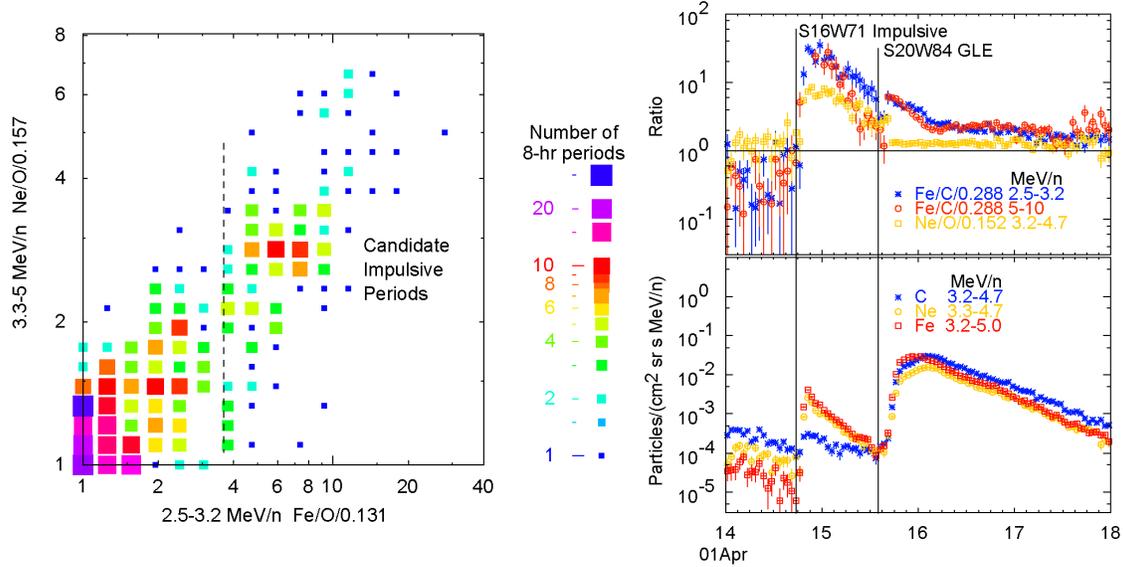

**Figure 2**. The left panel shows a histogram of Ne/O *vs.* Fe/O for all 8-hr periods with accuracy of 20% or better. The right panels show intensities (lower) and normalized abundance ratios (upper) during impulsive and gradual SEP events [11].

Note that Fe/C is enhanced relative to the coronal value in both events in the upper right panel of Figure 2, but much more so in the impulsive event. However, Ne/O is also strongly enhanced in the impulsive event, but not at all in the gradual event. While the event-averaged abundance of Fe/C or Fe/O may be adequate to distinguish acceleration mechanisms, Ne/O or $^3$He/$^4$He improves their resolution, as do the properties of associated CMEs and flares or type II or III radio bursts.

**4. Gradual SEP events and the FIP effect.**
Meyer [3] found that the element abundances in large SEP events consisted of a "mass-dependent" factor (actually dependent upon *A/Q*) organized by Fe/O and a simple dependence of the average behavior of the SEP/photospheric abundance ratios *vs.* the FIP of the element.

More recently, it has become clear that the *A/Q* dependence results from the differential transport of the ions from their point of acceleration to their point of observation. The scattering mean free path $\lambda$ of ions propagating outward along the magnetic field depends on a power of their magnetic rigidity. If we compare ions at the same velocity, as we always do, this becomes a power of *A/Q*. Frequently, the ions are scattering against self-generated Alfvén turbulence [13], however, even in the case of Kolmogorov turbulence $\lambda \sim (A/Q)^{1/3}$. Thus, Fe, with a larger value of *A/Q*, scatters less than C or O and the ions that arrive first are Fe-rich while those that arrive later are Fe-depleted or Fe-poor. As the shock wave expands across the spiral magnetic field, this temporal fractionation affects the longitude distribution as sketched in Figure 3. Thus, SEP events on the eastern flank of the shock are Fe-rich and those on the western flank are depleted in Fe/C as shown by the abundances in the sample events in Figure 3.

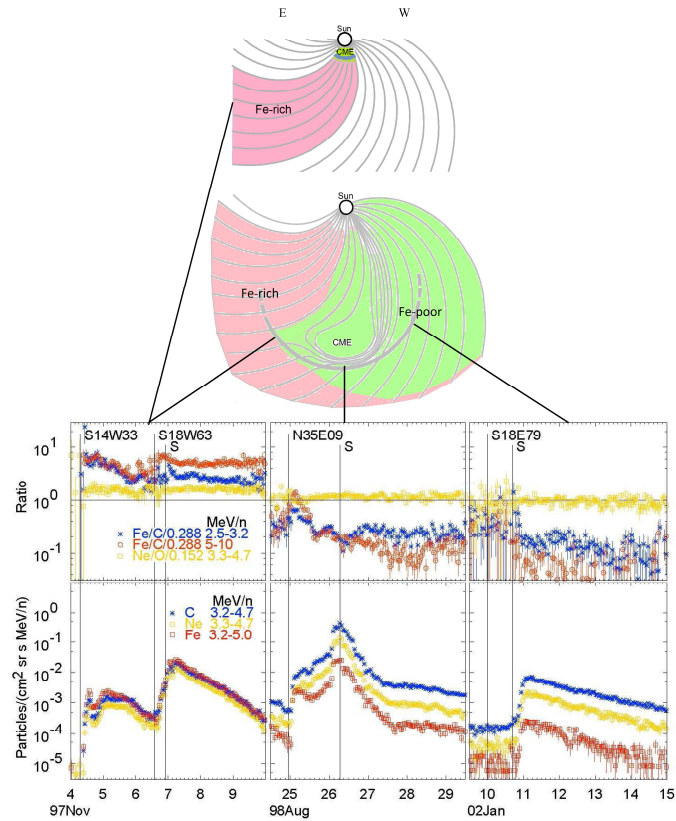

**Figure 3**. The lower panels show intensities (below) and relative abundance ratios (above) for four different SEP events viewing sources at the solar longitudes specified. S denotes a time of shock passage. Events viewed from the east of the source see Fe/C enhancements while those at central and western longitudes, relative to the source, see Fe/C depletions. Ne/O variations are minimal.

Averaging abundances over a large number of gradual SEP events tends to correct for these spatial variations. The average SEP abundances [10, 11], divided by two sources of photospheric abundances [14 and 15], normalized at O, are shown as a function of FIP in Figure 4 left and right panels.

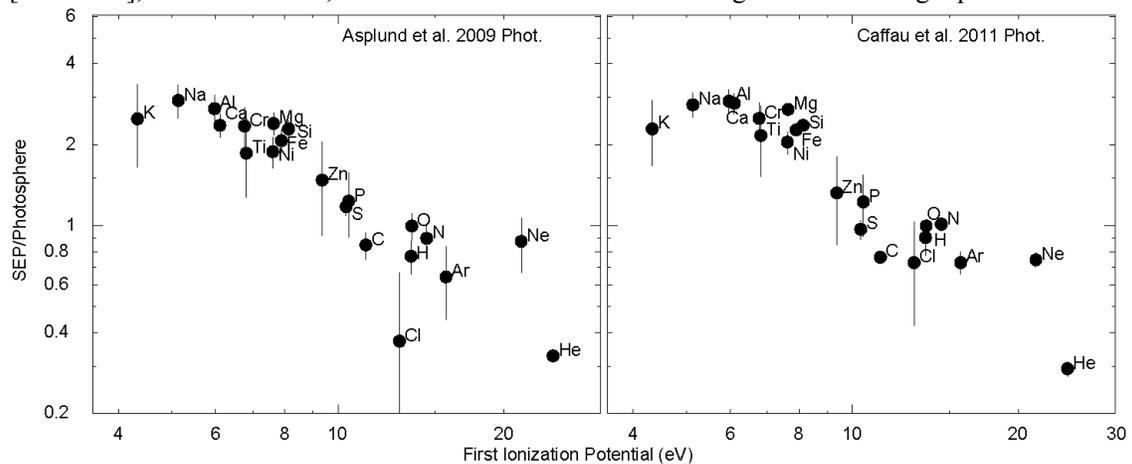

**Figure 4.** The ratio of the average SEP and photospheric abundances [14 and 15] is shown *vs.* FIP. Note that Fe and Mg (or Si), with different *A/Q* but similar FIP, fall together here.

Modern estimates of photospheric abundances involve models of the solar atmosphere. We have normalized the abundances in the left panel of Figure 4 to the photospheric abundances of Asplund *et al.* [14]. In a previous paper [11] we normalized to photospheric abundances of Caffau *et al.* [15] shown in the right panel (these photospheric abundances include no errors). Note that He is believed to be suppressed relative to the other high-FIP ions because of its unusually long ionization time.

The normalization of the FIP effect with respect to H is rather uncertain for SEP abundances alone, but is important for modeling the corona. The average behavior of the coronal abundances has been derived by Schmelz *et al.* [16] from SEP, solar wind, and spectral line abundances.

## 5. Abundance enhancements in impulsive SEP events

While the measurement of high-$Z$ ($Z \geq 34$) elements is fairly recent [6], attempts to associate the pattern of enhancements with the coronal temperature at the time of ion acceleration are not new [17]. Recent association of a large sample of impulsive SEP events with solar flares and CMEs by Reames, Cliver, and Kahler [18, 19] has renewed this study. If the enhancement of ions in magnetic-reconnection regions depends upon $A/Q$, and $Q$ depends upon coronal temperature $T$, then the abundance pattern depends upon $T$. The theoretical [20] dependence of $A/Q$ on $T$ is shown in Figure 5.

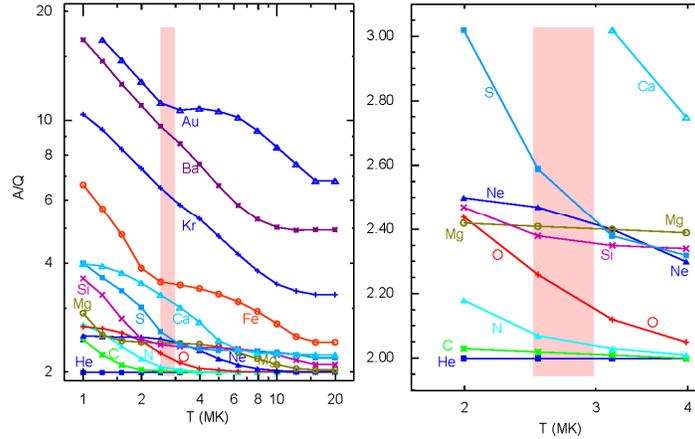

**Figure 5.** The theoretical [20] dependence of $A/Q$ *vs.* $T$ is shown for selected ions in the left panel and an expanded region is shown in the right panel [18].

If we assume values of $A/Q$ from the shaded region in Figure 5, we can plot the enhancement of elements averaged over 111 impulsive SEP events, relative to the corresponding abundance in gradual events, as a function of $A/Q$ in Figure 6 [18]. The fit in Figure 6 is the 3.6 power of $A/Q$.

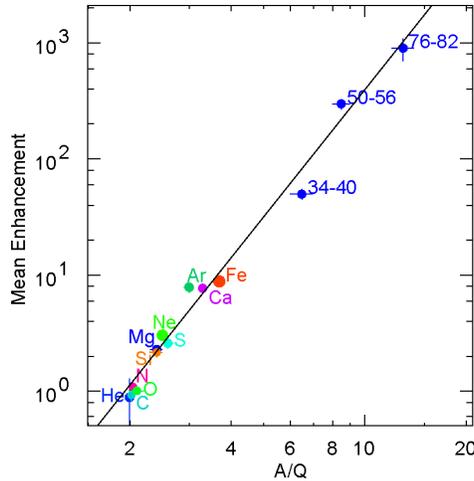

**Figure 6**. The average enhancement of elements in impulsive SEP events is shown *vs.* $A/Q$ [18].

Two aspects of the SEP abundance pattern that are important for selecting the coronal temperature are shown in Figure 7. In the left panel, all of the events with fast CMEs ($V>700$, red) and many others cluster around coronal values of C and He, but some are "C-poor" or "He-poor". This happens because $A/Q$ of O increases near 2.5 MK (see Figure 5). The right panel of Figure 5 shows that the enhancement in Ne exceeds that in Si in nearly all events. This inversion can be seen in the right panel of Figure 5 where the $A/Q$ value of Ne exceeds that of both Mg and Si in the 2–3 MK region.

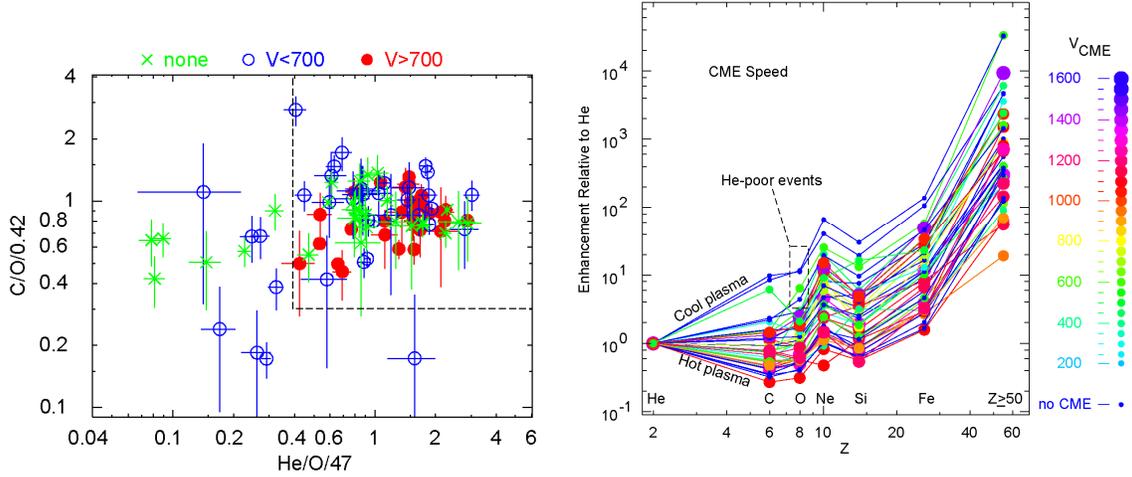

**Figure 7**. Normalized abundances in the left panel show some "C-poor" and "He-poor" events associated with slow CMEs [18]. The right panel shows enhancements that increase with Z have a discontinuity at Ne [18]. Fast CMEs tend to have more modest enhancements.

Thus, given that abundance enhancements increase as a power of $A/Q$ as expected theoretically [7], our choice of the shaded region of temperature in Figure 5 comes from three factors: 1) $A/Q$ of O begins to rise at low T, producing "He-poor" events, 2) $A/Q$ of Ne exceeds that of Mg and Si in the region, and 3) The power law extends smoothly to heavy elements for this region (Figure 6).

Statistically it is possible to determine the temperature and the power-law dependence, $(A/Q)^\alpha$, for each impulsive SEP event. If we select a temperature, that determines $A/Q$ for each species (Figure 5) and we can perform a least-squares, power-law fit to determine the power, $\alpha$, and the standard deviation of the fit, $\chi^2$. Sequencing through the temperatures, we select the one with the smallest value of $\chi^2$. When this is done, nearly all of the fits for the 111 SEP events are found to have temperatures of 2.5–3.2 MK [19]. Nearly all of the steep events with $\alpha \geq 6$ had $T=2.5$ MK [19].

Given $\alpha$ and $T$ for each event, we can study the event-to-event variations in these quantities and their dependence on the properties of the associated CME and flare [19]. Figure 8 shows the variation of $\alpha$ and of the mean values of $\alpha$ and $T$ with the 1–8 Å X-ray peak intensity of the associated flares. It is clear that the steepest values of $\alpha$ are associated with the He-poor events (circled), with the coolest temperatures, and with the smallest, B- and C-class X-ray flares. In fact, we have already seen from the right-hand panel of Figure 7 that the He-poor events have the steepest enhancements, not only at O, but continuing to much higher Z.

## 6. $^3$He/$^4$He in impulsive SEP events

$^3$He enhancements are just another property of impulsive SEP events. However $^3$He/$^4$He is known to be uncorrelated with Fe/C (*e.g.* [17]) and with other enhancements of heavy ions. Therefore it is appropriate to consider separately the source properties of $^3$He-rich events.

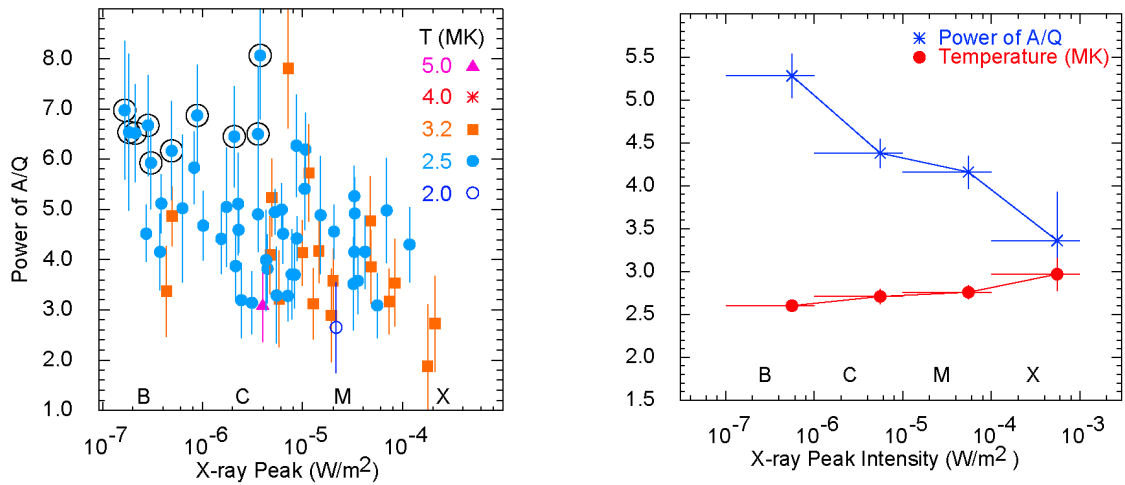

**Figure 8.** The left panel shows the abundance enhancement power of *A/Q*, *α*, *vs*. the 1–8 Å X-ray peak intensity of the associated flare for individual impulsive SEP events [19]. He-poor events are circled. The right panel shows the mean values of *α* and *T vs*. the X-ray peak intensity [19].

The left panels of Figure 9 compare the distribution of $\alpha$ (upper panel) and $^3$He/$^4$He (lower panel) for a plot of Ne/C *vs*. Fe/C. Note that $\alpha$ is extremely well correlated with Ne/C and Fe/C, as expected, but $^3$He/$^4$He in definitely not. Note also that a few events reach extreme enhancements of Fe/C≈100.

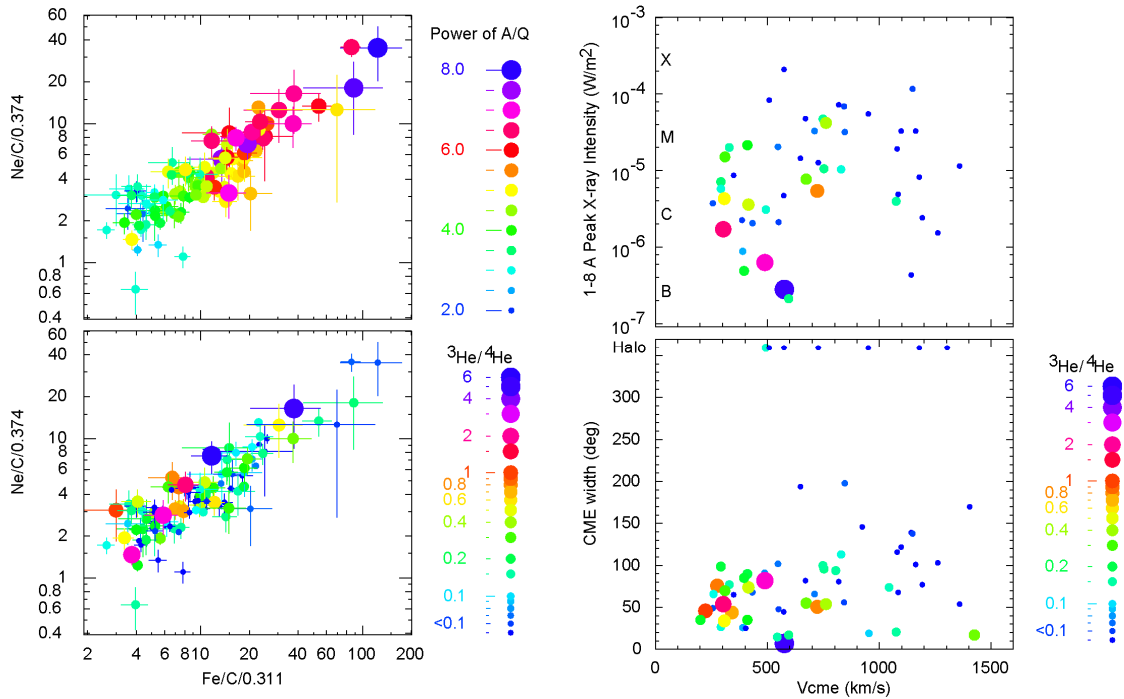

**Figure 9**. The left panels show values of $\alpha$ (above) and $^3$He/$^4$He (below) as varying symbols on a plot of Ne/C *vs*. Fe/C [19]. The right panels show the location of $^3$He/$^4$He enhancements of varying degree in plots of CME and flare properties [19].

The right panels of Figure 9 show where the most $^3$He-rich events fall on plots of CME width *vs.* CME speed (lower panel) and of X-ray peak intensity *vs.* CME speed (upper panel). The events that are the most $^3$He-rich come from narrower (<100°), slower ($V_{CME}$<800 km s$^{-1}$) CMEs. They also tend to be associated with small X-ray flares. The percentage of flares, by X-ray class, that have $^3$He/$^4$He>0.1 is B 79%, C 52%, M 32%, and X 0%.

## 7. Discussion

In the determination of abundances for gradual SEP events we have used only measurements in the energy region of ~2–12 MeV amu$^{-1}$. Much improved element resolution is possible at higher energies, and above 20 MeV amu$^{-1}$, isotope resolution is available for elements up to Ni [21]. However, above ~10 MeV amu$^{-1}$ energy spectra often break downward so as to produce a sharp rise or fall in abundance ratios [2, 12]. This can introduce large event-to-event variations that are difficult to correct. Abundances are also well measured below ~1 MeV amu$^{-1}$ [22]. However, these ions must propagate slowly through copious waves generated by all the ions of higher energies that have preceded them [23]. Fe scatters much less than C or O, so all events tend to look Fe-rich, and a much larger transport correction would be required at 0.5 MeV amu$^{-1}$ than at 5 MeV amu$^{-1}$.

The coronal abundances derived from gradual SEP events are now fairly well defined. Most sources (SEP, slow solar wind, and spectral lines) agree on the general behaviour of the coronal FIP effect, with the uncertainty in the *photospheric* abundances being nearly as large as that in coronal abundances [16]. It is also true that the different measurements come from differing spatial regions of the corona, some highly localized and some broadly averaged. Also, event-to-event abundance variations in the SEP events are not entirely statistical and are thus not entirely understood.

For the impulsive events, we are beginning to understand their general properties, but less so the variations. The power-law enhancements seem to come from magnetic reconnection [7] and the $^3$He enhancements from streaming electrons and wave-particle interactions [9]. The unusual enhancement in Ne is seen in both impulsive SEPs and in the broad γ-ray lines from flares [24]. This property helps define the coronal temperature of the ion source available for acceleration which is just below the active region core temperature of 3–4 MK [25]. Do the open-field sources of the SEPs occur on the cooler edges of active regions? It has been suggested that the edges of active regions are magnetically open to the solar wind [26].

The impulsive SEP properties are only loosely correlated with CME and flare parameters. Probably, the open-field region of the SEP acceleration is spatially distinct from most of the CME or flare. This may be especially true in larger events. Flare plasma can become extremely hot (>10 MK) in part because the heated plasma from the reconnection is contained on closed loops – in contrast to the escaping SEPs and plasma which carry energy away along open fields. The hottest plasma dominates the X-ray peak intensity. In essence, energy removal by the escaping SEPs may provide a kind of "evaporative cooling" for their local reconnection region.

It is important to note that SEPs *do not* come from hot (>10 MK) flare plasma. Either the SEPs come from a region that does not get hotter than ~3.2 MK or they are all accelerated early before the plasma is heated, perhaps both. The open-field line regions that allow SEPs to escape may not get hotter. In the closed field-line flaring regions that produce Ne-enhanced energetic ions that emit broad γ-ray lines [24], acceleration may occur early.

Why do the smallest flares and slowest CMEs often have the greatest SEP enhancements? Large SEP events may involve large regions which contain multiple smaller regions with differing enhancements; the combined result is an average enhancement. Small SEP events may sample a smaller more-uniform region, with a single well-defined enhancement, large or small, from the available distribution. From the left panel of Figure 8, for example, we see that the spread of α is greater for B- and C-class flares than for M- and X-class flares, which tend to be average.